White Paper for Plasma 2020 Decadal Survey

# Challenges and the next transformative steps in understanding plasma turbulence from the perspective of multi-spacecraft measurements

Li-Jen Chen
(Tel: 301-286-5358, Email: li-jen.chen@nasa.gov)
Heliophysics Science Division
NASA Goddard Space Flight Center

March 2019

Co-authors: Naoki Bessho[1], Jay A. Bookbinder[2], Damiano Caprioli[3], Melvin Goldstein[4], Hantao Ji[5], Lan K. Jian[4], Homa Karimabadi[6], Yuri Khotyaintsev[7], Kristopher G. Klein[8], Benoit Lavraud[9], William Matthaeus[10], Thomas E. Moore[4], Alessandro Retino[7], Owen W. Roberts[11], Vadim Roytershteyn[12], Conrad Schiff[4], Harlan Spence[13], Julia Stawarz[14], Jason TenBarge[5], Shan Wang[1]

[1]University of Maryland at College Park, [2]NASA Ames Research Center, [3]University of Chicago, [4]NASA Goddard Space Flight Center, [5]Princeton University, [6]Analytics Ventures, [7]Swedish Institute of Space Physics, Uppsala, Sweden, [8]University of Arizona, [9]Institut de Recherche en Astrophysique et Planétologie, Université de Toulouse,, Toulouse, France, [10]University of Delaware, [11]IWF, space research institute, Austrian academy of sciences, Graz , Austria [12]Space Science Institute, [13]University of New Hampshire, [14]Imperial College London, London, UK

We have become heavily reliant on electrical technologies, from power grids to GPS to wireless communication. Any disruption of these systems will have severe global consequences. A major natural hazard for such electrical disruption is caused by solar wind disturbances that have dramatic geospace impact. Estimates are that a solar storm of the magnitude of the 1859 Carrington Solar Superstorm would cause over $2 trillion in damage today [1]. In July 23, 2012, we had a near miss of a solar Superstorm that could have broken the record of largest such storms at Earth [2]. To enable pre-emptive measures, developing accurate space weather forecasts is urgent.

At the core of space weather forecasts is plasma physics, and kinetic turbulence, in particular. For example, the intense turbulence stirred up at the bow shock and foreshock have been shown to open up pathways for high velocity solar wind parcels to bypass the protective shield of the Earth's magnetosphere and create disturbances in the ionosphere and lower atmosphere [3-5].

A primary challenge in understanding kinetic turbulence and its global implications is its multi-scale nature, spanning from electron scales to macro scales of the magnetosphere. Current four-spacecraft missions with 3D formations, the Magnetospheric Multiscale (MMS) and Cluster, have made progress in our understanding of such turbulence. Yet the limitation of a fixed spacecraft formation size at a given time prohibits probing the multi-scale nature as well as the dynamical evolution of the phenomena. A transformative leap in our understanding of turbulence is expected with in-situ probes populating a 3D volume and forming multiple "n-hedrons (n>4)" in MHD to kinetic scales.

**Bow shock/foreshock/magnetosheath turbulence.**

An underlying connection between collisionless shocks, turbulence and magnetic reconnection leads to complex interactions in the geospace environment, as revealed by global kinetic simulations [3]. Ions reflected off the segment of the bow shock where the surface normal is quasi-parallel to the interplanetary magnetic field lead to self-generation of turbulence. This turbulence takes the forms of nonlinear waves that penetrate into the inner magnetosphere [6], and steepened magnetic structures resulting in plasma heating, particle acceleration and reconnection [7,8]. At the shock, the turbulence causes the formation of high speed jets that penetrate the bow shock and bombard the magnetopause a few times per hour leading to dayside reconnection [5]. These large-scale structures are generated by kinetic processes. Turbulence also gives rise to intense current sheets, which can reconnect and form magnetic islands, both at the shock transition layer and extending deep into the downstream magnetosheath [3,9]. Electron scale current sheets form at the shock transition layer and reconnect, as recently predicted by fully kinetic simulations (Figure 2, left panel), and observed by the MMS spacecraft [10,11].

The magnetic field spectrum from an example shock containing ion-scale reconnecting current sheets (right panel of Figure 2, blue curve) presents a broad peak of enhanced magnetic energy in the vicinity of the ion inertial length ($d_i$) scale (marked by a vertical solid line and labeled as $<f_{di}>$). The foreshock (characterized by the presence of ions reflected off the bow shock and intense magnetic fluctuations with typical dB/B >> 1) immediately upstream of the example bow shock exhibits multiple magnetic spectral peaks at ion and electron scales. The shown spectral features indicate multi-scale energy injections. In the example, the dynamics of coherent structures in the form of $d_i$-scale current sheets plays a critical role in dissipation. Downstream from quasi-parallel shocks in the magnetosheath, intense current sheets both reconnecting [9,12] and non-reconnecting are abundant.

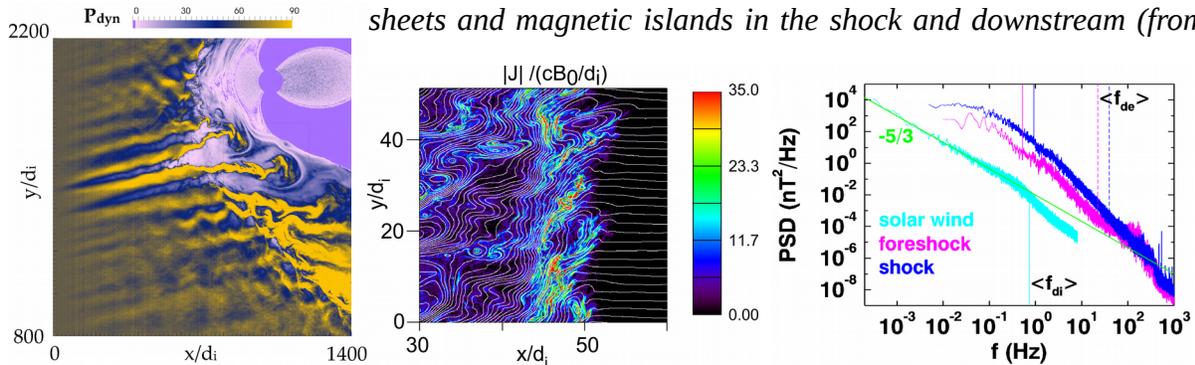

*Figure 1: (left) High velocity jets, due to turbulence at the shock, can reach the magnetopause, triggering reconnection and space weather effects [3,5]. (Middle) Intense reconnecting current sheets and magnetic islands in the shock and downstream (from a fully kinetic simulation). (Right) Magnetic field spectra from the foreshock and bow shock populated by reconnecting current sheets [10]. The strong magnetic turbulence is convected into the magnetosheath where kinetic scale current sheets are abundant.*

**Major achievements and challenges from the MMS and CLUSTER missions**

The major achievements of the MMS mission on shock/foreshock/magnetosheath turbulence are : (1) Reconnecting current sheets were discovered in the turbulent shock transition region [10,11], and downstream magnetosheath [12]. (2) In magnetosheath turbulence, structures with high intermittency are regions of significant electron heating with typical structures showing features consistent with magnetic reconnection [13]. (3) The energy partition between magnetic fields and plasma flows was examined for the magnetosheath, where the kinetic scale energy partition was found to deviate from that in the MHD scale and the electron flow energy dominates at the electron scale [14]. (4) The wave modes in the magnetosheath above the ion scale were identified using the 4-spacecraft k-filtering technique [14]. For MMS, the multi-spacecraft technique cannot resolve smaller wavelengths than the inter-spacecraft separation (mostly electron scales) and the spatial scale of fast propagating structures due to the combined limitation of the small spacecraft separation and time resolution.

The major achievements of the Cluster mission on shock/foreshock/magnetosheath turbulence are : (1) the dissipation due to reconnecting current sheets is estimated to be about two orders of magnitude higher than that due to wave damping [15,16]. (2) The anisotropy of solar wind turbulence between the ion and electron gyroscales is measured by simultaneously probing a variety of directions (along pairs of the spacecraft) relative to the local magnetic field based on the four Cluster measurements and found to highly elongated along the magnetic field [17]. The full 3D anisortropy was also measured by the k-filtering method at proton scales [18].

(3) the energy injection at the foreshock is determined to be at ion scales, and instabilities responsible for the waves are identified [19].

Both missions lack cross-scale capabilities that can only come from having more spacecraft with multi-scale separations at any time. Having N spacecraft provides a maximum of N!/(4!(N-4)!) tetrahedra. The analysis capability quickly increases with an increasing number of measurement points. For example, having four spacecraft yields at most one tetrahedron, while 11 spacecraft could provide 330 tetrahedra!

One of the common challenges for studying self-generated turbulence is to obtain spatial scales reliably. Along the propagation direction, the spacecraft separation needs to be large enough to observe significant variations of the fields and smaller than the minimum wave length for both k-filtering (a technique enabling mapping from the frequency to wave vector space) [20] and timing analysis. The scale size along directions other than the convection flow is mostly unknown. In addition, the existing multi-spacecraft analysis methods assume stationarity as the structure propagates through the spacecraft cluster. The multi-spacecraft analysis for spatial gradients requires approximations to tetrahedrons or higher order (n-hedron, n>4) shapes of the spacecraft constellation.

**Future steps**

Simultaneous measurements from multiple spacecraft claiming a 3D volume and simultaneously covering the MHD to kinetic scales will provide a transformative step forward in multi-scale *in-situ* observations. Such a mission is particularly timely and will enable a closer tie with new multi-scale laboratory experiments such as FLARES [21] as well as upcoming Petascale 3D global kinetic simulations.